\documentclass[11pt,preprint]{aastex}

\begin{document}

%\received{20 Aug 2008}
%\accepted{11 Oct 2008}
%\journalid{number}{date month year}
%\articleid{number}{number}
%\slugcomment{submitted to: {\it AJ}}
%\slugcomment{accepted to: {\it }}

\title{The effect of the dust size distribution on asteroid polarization}
\author{Joseph Masiero\altaffilmark{1}, Christine Hartzell\altaffilmark{2}, Daniel J. Scheeres\altaffilmark{2}}

\altaffiltext{1}{Institute for Astronomy, University of Hawaii, 2680 Woodlawn Dr, Honolulu, HI 96822, {\it masiero@ifa.hawaii.edu}}
\altaffiltext{2}{Department of Aerospace Engineering Sciences, University of Colorado at Boulder, 429 UCB, Boulder, CO 80309-0429, {\it christine.hartzell, scheeres@colorado.edu}}

\begin{abstract}

We have developed a theoretical description of how of an asteroid's
polarization-phase curve will be affected by the removal of the dust from the
surface due to a size-dependent phenomenon such as radiation pressure-driven
escape of levitated particles.  We test our calculations against new
observations of four small ($D\approx1~$km) near-Earth asteroids [(85236),
(142348), (162900) and 2006 SZ$_{217}$] obtained with the Dual Beam Imaging
Polarimeter on the University of Hawaii's 2.2 m telescope, as well as
previous observations of (25143) Itokawa and (433) Eros.  We find that the
polarization of the light reflected from an asteroid is controlled by the
mineralogical and chemical composition of the surface and is independent of
dust particle.  The relation between the slope of the polarization-phase
curve beyond the inversion angle and the albedo of an asteroid is thus
independent of the surface regolith size distribution and is valid for both
Main Belt and Near-Earth asteroids.

\end{abstract}


\section{Introduction}

The scattering of light can induce a polarization to an otherwise unpolarized
photon flux.  In the case of asteroids, the dust and regolith on the surface
act to instill a linear polarization on the reflected sunlight.  The degree
of this polarization changes with phase angle, albedo, spectral type, and
surface composition \citep{muinonenAIII}.  While single-particle scattering
will always result in polarization perpendicular to the scattering plane
(``positive'' polarization), polarization in the scattering plane
(``negative'' polarization) is observed for very small phase angles for a
wide range of objects \citep[e.g.][etc.]{dz79}.  \citet{shkuratov94} present
a review of possible physical causes of this effect, the most promising of
which are coherent backscattering models \citep[e.g.][]{muinonen89} which
also can describe the photometric opposition effect.

In this paper, we determine the effects of different size distributions of
surface dust on the observed relationship between the polarization of the
reflected light and the phase angle of the asteroid (that is, the
Sun-asteroid-Earth angle).  A variety of processes can cause dust loss on
asteroids.  Electrostatic charging of surface dust particles by Solar
radiation, for example, can levitate smaller particles meters above the
surface \citep[][and reference therein]{colwell05}.  For small bodies these
effects will result in the preferential loss of the smallest dust grains,
effectively truncating the original dust size distribution.  Here we test six
cases of small near-Earth asteroids (NEAs), for which the dust loss should be
more pronounced, and measure the effect of dust loss on the
polarization-phase relation.

\section{Grain size limits}
\label{grainsize}

The surfaces of most atmosphereless bodies in the inner Solar system are
composed of a loose regolith ranging in size from dust to boulders.  On the
moon the smallest particles are dust down to $0.02~\mu$m in size
\citep{taylor09} while for smaller asteroids such as (25143) Itokawa the
regolith appears significantly coarser, with an apparent lack of anything
smaller than $1~$mm \citep{fujiwara06}.  With a surface gravity significantly
higher than all but the few largest asteroids, the dust distribution on the
moon sets a strong lower-limit on the dust particle sizes we will consider in
this study.  As we will show in \S~\ref{pol}, the polarization-phase relation
is insensitive to the upper-limit of the grain size distribution as long as
that limit is $>10~\mu$m.

Dust particles lofted from the surface of an asteroid by micro-meteoroid
bombardment or electrostatic levitation are immediately subject to Solar
radiation pressure effects.  This pressure can put the particle into orbit
around the parent body or allow the particle to escape into a heliocentric
orbit.  A lofted particle's motion can be described by Eq~\ref{eq:eom} which
includes the effects of both gravity due to the asteroid (where $G$ is the
universal gravitational constant, $M$ the mass of the asteroid, and
$\mathbf{r}$ the position vector of the particle in $x,y,z$ space) and Solar
radiation pressure ($g_s$). The radiation pressure is assumed to act in the
anti-Sun direction, $\mathbf{\hat{x}}$, with no time dependency (i.e. the
x-axis corresponds to the Sun-asteroid line and the orbit of the asteroid
around the Sun is not included in this approximation). The Solar radiation
pressure is described by Eq~\ref{eq:g}, where $p_v$ is the albedo of the
particle (assumed to be zero initially), $C_s$ is a constant of Solar
radiation ($1 \times 10^7~g~km^3/(s^2~cm^2)$), $B$ is the mass-to-area ratio
of the particle, and $R$ is the distance to the Sun.  The motion of the
particle can then be described by the potential given in Eq~\ref{eq:U}.

\begin{eqnarray}
\label{eq:eom}
\ddot{\mathbf{r}}&=& -G\frac{M}{r^3} \mathbf{r}+ g_s \mathbf{\hat{x}}\\
\label{eq:g}
g_s &=& \frac{(1+p_v) C_s}{B R^2}\\
\label{eq:U}
U&=&G\frac{M}{r}+g_s \left(\mathbf{\hat{x}} \cdot \mathbf{r} \right)
\end{eqnarray}

A mathematical analysis of the orbital dynamics of test particles at an
asteroid has been carried out previously by several authors.
\citet{dankowicz94} developed a precise limit on semi-major axis of an
orbiting particle under the assumption that the asteroid was stationary.
\citet{scheeres02} incorporated the motion of the asteroid about the Sun and
developed a sufficient condition on semi-major axis for a particle (in this
case a spacecraft) to be trapped in orbit about a small body.
\citet{byram09} combined elements of these two earlier studies to derive more
precise limits on semi-major axis of an orbiting test particle accounting for
its motion.  For purposes of this paper the analysis in \citet{byram09} is
too detailed to be summarized simply, and thus we will only restate the
results of these earlier papers.

From the analysis by \citet{dankowicz94} we find that a simple limit exists
for the semi-major axis of a lofted particle that generally ensures escape
from the small body if exceeded.  Mapped into our notation we find that
escape will occur if the semi-major axis is larger than $a_{escape}$:
\begin{eqnarray}
	a_{escape} & = & \frac{\sqrt{3}}{4} \sqrt{G\frac{M}{g_s}} \label{eq:aescape}\\
	& = & \frac{\sqrt{3}}{4} \sqrt{G\frac{M~B}{C_s}} R 
\end{eqnarray}
assuming an albedo of $p_v=0$.  Using a very different approach,
\citet{scheeres02} developed a sufficient condition for stability about a
small body and found that escape is not immediately possible if the
semi-major axis is less than $a_{min}$:
\begin{eqnarray}
	a_{min} & = & \frac{1}{4} \sqrt{G\frac{M~B}{C_s}} R
\end{eqnarray}
In the general problem it is possible for the semi-major axis to evolve, but
we ignore this effect for this derivation.  We note that these limiting
semi-major axes scale linearly with the small body distance from the Sun.
Thus, escape is most likely to occur at perihelion.

In order to calculate the mass-to-area ratio ($B$) of a particle, we assume
that it is spherical with a circular reflective surface area, giving
$B=\frac{2}{3}~\rho_d~d_d$, where $\rho_d$ is the dust density and $d_d$ is
the diameter of the dust particle.  Note that these assumptions are
conservative as spheres maximize the volume-to-area ratio, and thus these
calculations provide a lower limit to the cross-sectional surface area of
particles escaping.  Additionally, we are ignoring any timescale for particle
loss once in orbit, assuming that sufficient time has passed to clear all
particles in the perihelion size-limit.  From this we can expand
Eq~\ref{eq:g} to find:
\begin{eqnarray}
\label{eq:gexpanded}
g_s= \frac{3}{2} \frac{(1+p_v) C_s}{\rho_d~d_d~R^2}
\end{eqnarray}

Substituting Eq~\ref{eq:gexpanded} into Eq~\ref{eq:aescape} and assuming that
the particle of interest rests on the asteroid's surface ($2a=D_{ast}/2$,
where $D_{ast}$ is the asteroid's diameter), we can develop an expression
that relates the minimum particle size expected to be found on an asteroid's
surface to the physical characteristics of the asteroid:
\begin{equation}
d_d \leq \frac{3C_s}{\pi G} \frac{p_v+1}{D_{ast} \rho_{ast} \rho_d R^2 }
\label{eq:rd}
\end{equation}

\begin{figure}
\centering
\includegraphics[angle=-90,scale=0.6]{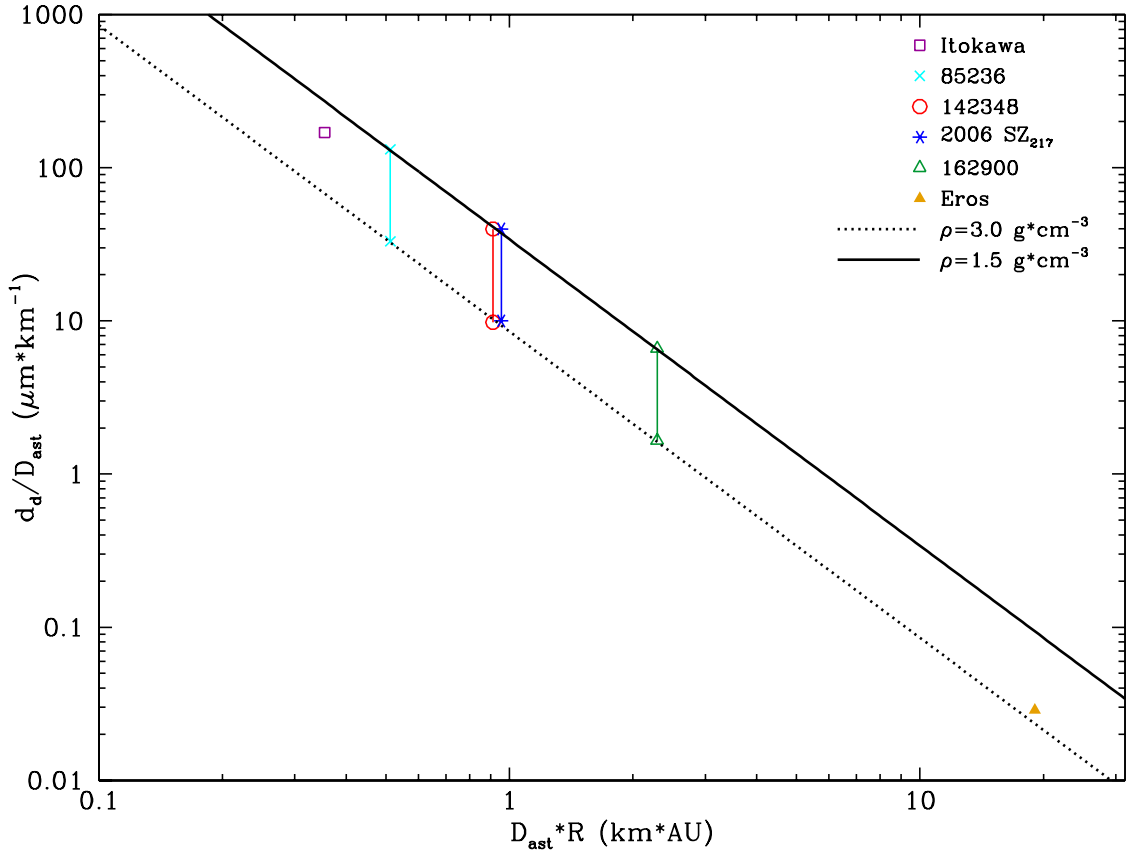}
\caption{Ratio of the minimum dust particle size on the surface to the diameter of the asteroid, as related to the asteroid's diameter times the distance to the Sun.  The solid and dotted lines show the relationship for constant densities of $1.5$ and $3.0~$g cm$^{-3}$, respectively, when assuming an albedo of $0.2$.   The points indicate the positions of NEAs considered in this study with known densities, while the ranges represent NEAs modeled with densities from $1.5$ to $3.0~$g cm$^{-3}$.  Deviations of the calculated points from the lines of constant density are due to slight differences in the measured albedos.}
\label{fig.trend}
\end{figure}

Equation~\ref{eq:rd} shows that as the size of the central body decreases or
as perihelion distance becomes smaller, the minimum surface particle size
increases.  This can be seen in Fig~\ref{fig.trend}, which shows with
thick lines the relationship given in Eq~\ref{eq:rd} (normalized by the
central body diameter) assuming the central body and the dust particle have
the same density and an albedo of $p_v=0.2$.  We note that dust should have
densities up to a factor of $2$ more than the bulk asteroid density, however
this is mitigated by the assumption of spherical shapes for particle size.
For asteroids spinning near their disruption rate we can assume that
$a=D_{ast}/2$, resulting in a minimum particle size that is a factor of four
larger than predicted by Eq~\ref{eq:rd}.

Table~\ref{tab:examples} shows the particle size limit for (433) Eros and
Itokawa, ignoring any rotation.  Note that the minimum particle size
is inversely related to the asteroid density, diameter and distance from the
Sun.  If we consider the migration of Itokawa from the Main Belt to its
current location, we see that as the asteroid approaches the Sun,
increasingly larger particles are able to escape.  If the polarization of
light reflected from an asteroid is dependent on the size of the particles on
the surface we would expect NEAs to show changes in their polarization-phase
curve as a function of the predicted particle size.

\section{Polarization Effects}
\label{pol}

The coherent backscatter mechanism, a second order light-scattering effect,
has been shown to be the dominant cause of both the observed polarization
signature of light reflected from asteroids at small phase angles as well as
the photometric opposition effect \citep{muinonenAIII}.  As phase angle
increases this transitions to a regime dominated by first-order
single-particle scattering.  Using numerical techniques to calculate the
scattering of individual surface elements, \citet{muinonen02} find that the
polarization of the light can be approximated as:
\begin{eqnarray}
\label{eq.P1d}
P \sim \frac{\alpha^2}{2 n} - \left(\frac{n-1}{n+1}\right)^2 \frac{(k~d~\alpha)^2}{2~[1+(k~d~\alpha)^2]} 
\end{eqnarray}
where $P$ is the percent polarization, $\alpha$ is the phase angle in
radians, $n$ is the index of refraction, $k=\frac{2\pi}{\lambda}$ is the wave
number, and $d$ is the scatter separation distance.  This assumes a single
spacing for all scattering particles.  Other recent models
\citep[e.g.][]{boehnhardt04,bagnulo06} have assumed two discrete scattering
components co-mixed at different weightings to fit observed polarization
phase curves.

In this work we will instead consider a continuous distribution of grain
spacings.  The distribution of voids in a silicate glass follows an
approximate power-law falloff for sizes beyond some critical peak void size
\citep{malavasi06}.  We begin by assuming that the spacing between dust on an
asteroid's surface also follows a power law, and that the voids have the same
minimum and maximum characteristic sizes as the dust.  This implies that
chemical and mineralogical effects are fully incorporated into the size of
the constituent grains, which we discuss further in \S~\ref{conc}.

\citet{bottke05} show that for their model NEA population the size
distribution is fit by the power law $N \propto D^{-2.5}$ for sizes from
$100~$m to $<10~$cm.  Using a large-area camera network, \citet{halliday96}
find that the mass distribution of fireballs ($0.1 \le M \le 12~$kg) entering
the Earth's atmosphere has an exponential power between $-0.5$ and $-1$,
translating to a size distribution power between $-1.5$ to $-3$.  We focus on
this range of size distribution powers for our simulations.

By integrating Eq~\ref{eq.P1d} over all grain sizes from $d_{min}$ to
$d_{max}$ we obtain:
\begin{eqnarray}
\label{eq.Palld}
P = A \int_{d_{min}}^{d_{max}}\frac{\alpha^2}{2 n} - \left(\frac{n-1}{n+1}\right)^2 \frac{(k~\bar{d}~\alpha)^2}{2~[1+(k~\bar{d}~\alpha)^2]} \bar{d}^p\,d\bar{d}
\end{eqnarray}
where $A$ is a variable used to fit the depth of the negative polarization,
$p$ is the exponential power of the grain-size distribution, and $d_{min}$
and $d_{max}$ are minimum and maximum sizes of the grain-size distribution in
microns, respectively.

If we consider only a single wavelength (in this case $\lambda \approx
600~$nm, meaning $k = 10~\mu$m$^{-1}$) we find that the arbitrary variable
$A$ reduces to a simple expression of $d_{min}$, giving:
\begin{eqnarray}
\label{eq.PalldClean}
P = 1.2~d_{min} \int_{d_{min}}^{d_{max}}\frac{\alpha^2}{2 n} - \left(\frac{n-1}{n+1}\right)^2 \frac{(10~\bar{d}~\alpha)^2}{2~[1+(10~\bar{d}~\alpha)^2]} \bar{d}^p\,d\bar{d}
\end{eqnarray}

\begin{figure}
\centering
\includegraphics[angle=-90,scale=0.6]{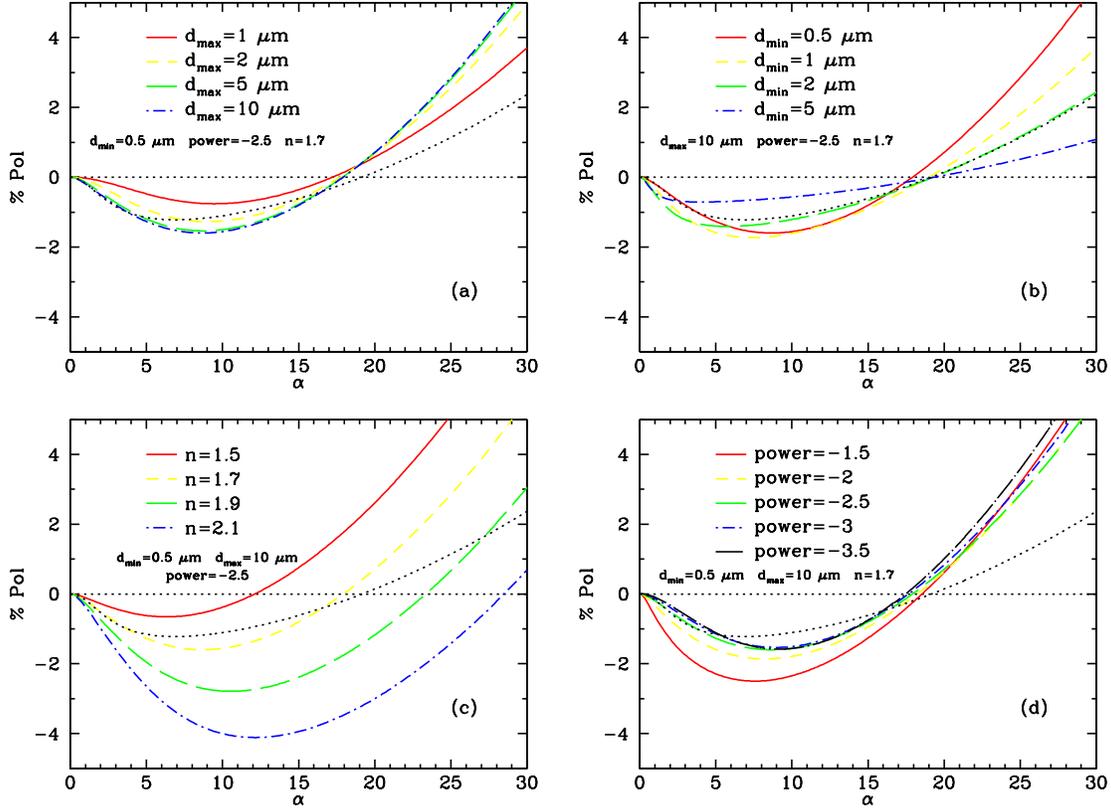}
\protect\caption{
Simulated polarization-phase curves for variations in the value of (a) the maximum grain size, (b) the minimum grain size, (c) the index of refraction, and (d) the power of the size distribution.  The curves show the relation between percent of incident light polarized, referenced to the normal to the scattering plane, and the Solar phase angle ($\alpha$) measured in degrees.  The dotted curve in each box shows the polarization described by Eq~\ref{eq.P1d} for $n=1.7$, $(k~d)=20$, and a normalizing constant of $50$: a good approximation for the polarization from a typical S-type Main Belt asteroid.  
}
\label{fig.model}
\end{figure}

Figure~\ref{fig.model}a shows the change in the polarization-phase relation for
fixed $n$, $d_{min}$ and $p$.  As $d_{max}$ increases its effect on the curve
diminishes, meaning that the polarization is insensitive to the largest
particle size for general distributions.  For all other simulations we fix
the maximum particle size at $10~\mu$m to limit computation time.  In
Fig~\ref{fig.model}b we plot the effects of a variable $d_{min}$.  As a
normalizing agent, $d_{min}$ has a strong effect on the slope of the relation
beyond the inversion angle (the point at which the polarization-phase curve
recovers to zero), but also on the shape and depth of the negative branch of
polarization.

A variable index of refraction most significantly affects the location of the
inversion angle $\alpha_0$.  As shown in Fig~\ref{fig.model}c small changes in
the index of refraction result in major changes to the inversion angle and
depth of negative polarization, with only minor effects on the slope or shape
of the negative branch.  Finally, the effect of variations in the power of
the size distribution are shown in Fig~\ref{fig.model}d.  For the ranges of
powers considered there are only minor changes to the slope and inversion
angle, and only in the extreme case is there a significant change to the
negative polarization branch.

From Eq~\ref{eq.PalldClean} we can calculate the properties of the
polarization-phase relation in terms of physical constants only.  We find
that the slope for $\alpha > \alpha_0$ is described by:
\begin{eqnarray}
\label{eq.slope}
h = \frac{-1.275}{n \left(p+1\right)}~d_{min}~\left({d_{min}}^{p+1} - {d_{max}}^{p+1}\right)
\end{eqnarray}
where $h$ is the slope (in percent polarization per degree).  As $d_{max}$
becomes large ($>>10~\mu$m) it's effect on $h$ becomes negligible.

Similarly, for $n \ge 1.5$ and $d_{max} \ge 10~\mu$m we can describe the
location of the inversion angle approximately as:
\begin{eqnarray}
\label{eq.a0}
\alpha_0 \approx \sqrt{n \left(\frac{n-1}{n+1}\right)^2 - \frac{1}{2~\left(10~d_{min}\right)^2} }
\end{eqnarray}
where $\alpha_0$ is the inversion angle in radians.

\citet{cellino99} derive updated constants for the relationship between
polarization slope $h$ and geometric albedo $p_v$:
\begin{eqnarray}
\label{eq.albedo}
\log{p_v} = -1.118 \log h - 1.779
\end{eqnarray}
Using Eq~\ref{eq.slope} and assuming that $d_{max}>>10~\mu$m we can now
describe the geometric albedo of an asteroid as:
\begin{eqnarray}
\label{eq.albedoNew}
\log{p_v} = 1.118 \log{\left(n~(-p-1)~{d_{min}}^{-p-2} \right)} - 1.896
\end{eqnarray}

\section{Observations}
\label{obs}

In order to test the effect of dust depletion on the polarization of small
NEAs, we conducted observations of four targets over a five week period at
the end of 2008 using the Dual-Beam Imaging Polarimeter (DBIP) on the
University of Hawaii's 2.2~m telescope \citep{dbip}.  DBIP is designed to
measure both linear and circular polarizations of asteroids and other point
sources at magnitudes $10 \le V \le 17$ with errors less than $0.1\%$
\citep{dbip2}.

Targets were chosen from all small NEAs ($D\approx1~$km) that would be
brighter than $V=18$ over the observing window.  Four asteroids had
polarimetric measurements of sufficient quality to compare with the theory
discussed above.  Asteroids (85236), (142348), (162900) and 2006 SZ$_{217}$
were measured at a range of phase angles sufficient to determine both
$\alpha_0$ and $h$.  Table~\ref{tab.obs} gives the target name, UT date of
observation, apparent $V$ magnitude, exposure time, number of exposures,
Solar phase angle $\alpha$, measured linear polarization, and angle of
polarization referenced to the vector orthogonal to the Sun-object-Earth
scattering plane.

In Table~\ref{tab.dust} we present for each asteroid the fitted inversion
angle and slope (considering only phase angles $\alpha \ge 15^\circ$) as well
as perihelion distance (q), absolute magnitude (H$_v$), albedo (as calculated
from Eq~\ref{eq.albedo}), diameter (D), $n$ and $d_{min}$ calculated from
Eqs~\ref{eq.slope} and \ref{eq.a0} (assuming $p=-3$ and $d_{max}=10$), and
the range of predicted minimum particle sizes remaining on the surface
($d_{pred}$) depending on the assumed density (between $1.5$ and
$3.0~$g~cm$^{-3}$) following \S~\ref{grainsize}.  Included in the table are
the same values for Eros \citep{zgEros} and Itokawa \citep{cellinoItokawa},
two NEAs that have been extensively studied both remotely as well as
\textit{in situ} by spacecraft visits.  A measured density for each of these
objects \citep{erosDensity,itokawaDensity} allows us to make a singular
prediction for minimum surface particle size.  The $d_{pred}$ range or value
for each of the six asteroids is shown in Fig~\ref{fig.trend}.  Scaling the
levitation calculations from \citet{colwell05} to smaller asteroids we find
our predicted minimum particle sizes are within the range of particle sizes
that are capable of being levitated, assuming the particle's surface electric
potential is equal to that of the asteroid's surface.

\begin{figure}
\centering
\includegraphics[scale=0.6]{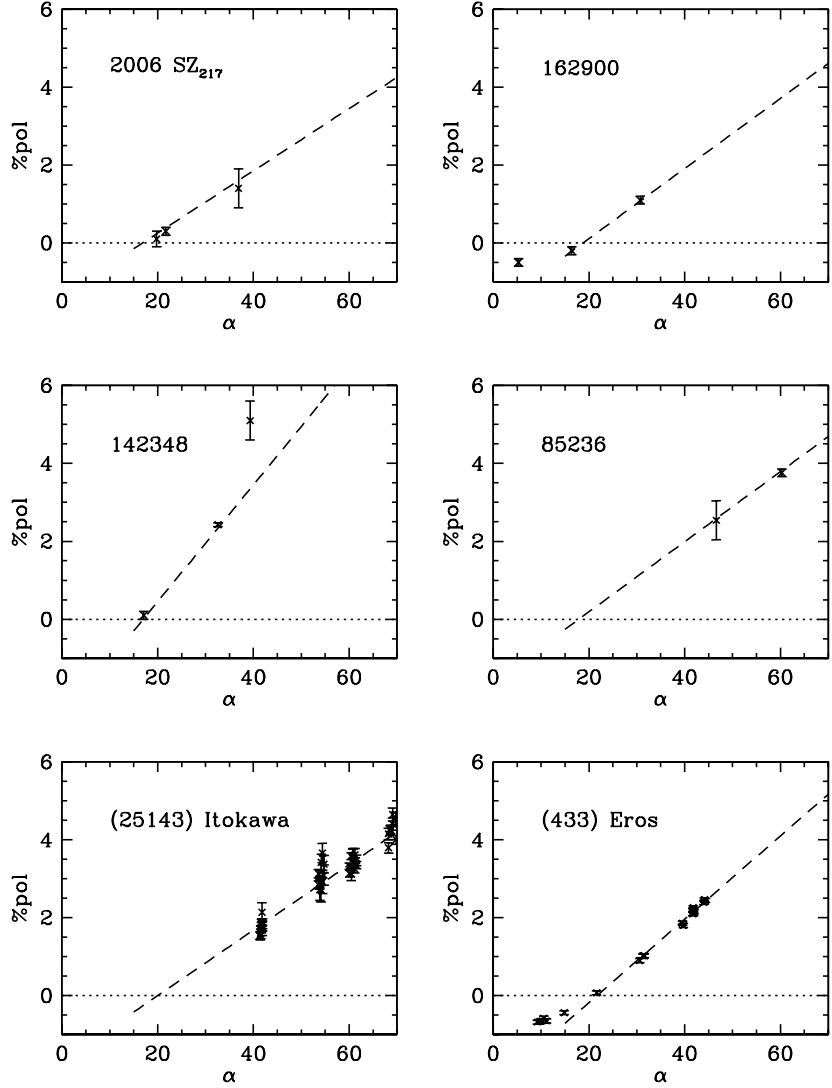}
\protect\caption{
Observations from Table~\ref{tab.obs}, as well as the best fitting linear polarization-phase relation for phase angles $\alpha \ge 15^\circ$ (dashed line).  Fitted slopes and inversion angles are given in Table~\ref{tab.dust}.  Note that 2006 SZ$_{217}$ and (85236) are under-constrained.  Data for Itokawa is from \citet{cellinoItokawa} while data for Eros is from \citet{zgEros}.
}
\label{fig.polplot}
\end{figure}

Figure~\ref{fig.polplot} shows our observations and fits for each asteroid,
as well as for the literature data for Itokawa and Eros.  Note that the fits
for 2006 SZ$_{217}$ and (85236) are under-constrained, thus values for the
slope have large errors and the inversion angles are only approximate.  In
all cases, the polarization-phase relation beyond $\alpha_0$ was assumed to be
strictly linear.

\section{Results and Discussion}

Calculations of the dust retention for the NEAs investigated predict over two
orders of magnitude in variation of the minimum particle size.  We find no
correlation in our data between the $d_{min}$ calculated from our measured
slope and the predicted minimum particle size.  If the individual dust
particles were unique scattering elements we would expect the opposite to be
true.  Thus the physical size of the dust cannot be the primary cause of
polarization or even a significant contributor.

In fact, (85236), (162900), 2006 SZ$_{217}$, Itokawa and Eros all show slopes
and inversion angles consistent with those measured for large S-type Main
Belt asteroids (MBAs), while (142348) shows polarization similar to what is
seen for large C-type MBAs \citep[see, e.g][]{muinonenAIII}.  From this we
infer that the polarization of light reflected from an asteroid is
constrained primarily by the way the local mineralogy and chemistry dictate
the formation of spaces in the mineral matrix.  The calculated value of
$d_{min}$ then represents not the size of the smallest particle on the
surface, but instead a characteristic minimum size of mineralogical voids.
For asteroids with similar spectral signatures, and thus similar surface
chemistries, we expect to see correspondingly similar polarization
properties.  Asteroid polarization-phase curves should show no variation with
changing regolith conditions for asteroids of the same composition.

The polarization-albedo relation (Eq~\ref{eq.albedo}) then does not depend on
surface regolith size distribution and so is identical for NEAs and MBAs.
Variations in surface regolith and observing geometries between NEAs and MBAs
require thermal models tailored for each population \citep{harrisNEATM,
wolters09} which in turn affects infrared albedo determinations.
Polarimetrically determined albedos are therefore immune to this complication
and can be robustly applied to any asteroid regardless of location in the
Solar system.

\section{Conclusions}
\label{conc}

Through modeling of the effects of a distribution of spacings between
scattering elements on the observed polarization-phase relation, we have
shown that the relation is independent of the size of the particles on the
surface.  Instead, the slope of the relation beyond the inversion angle (and
thus the spacing parameter $d$) remains constant for a range of objects with
dust-retention sizes spanning two orders of magnitude.  The
polarization-phase relation is therefore consistent between MBAs and NEAs of
the same spectral type.  An asteroid's surface chemistry and mineralogy are
the dominant cause of the polarization-phase relation, and this relation is
equally applicable for both NEAs and MBAs.

\section*{Acknowledgments}

We would like to thank Robert Jedicke and the anonymous referee for their
comments on the paper that greatly improved the text.  J.M. was supported
under NASA PAST grant NNG06GI46G.  D.J.S. acknowledges support from NASA's
Discovery Data Analysis Program.  The authors wish to recognize and
acknowledge the very significant cultural role and reverence that the summit
on Mauna Kea has always had within the indigenous Hawaiian community. We are
most fortunate to have the opportunity to conduct observations from this
sacred mountain.

\newpage

\begin{deluxetable}{lcccc}
\tablenum{1}
\tabletypesize{\footnotesize}
\tablecaption{Calculated minimum particle sizes from the dynamical model}
\tablewidth{0pt}
\tablehead{
\colhead{}   &
\colhead{Eros}   &
\colhead{Itokawa}   &
\colhead{Itokawa}   &
\colhead{Itokawa}   \\
\colhead{}   &
\colhead{Perihelion}   &
\colhead{Main Belt}   &
\colhead{Aphelion}   &
\colhead{Perihelion}  
}
\startdata
Asteroid Diameter (km) & 16.84 & 0.33 & 0.33 & 0.33 \\
Distance from Sun (AU) & 1.133 & 2.500$^a$ & 1.695 & 0.953 \\
Density ($g/cm^3$) & 2.7 & 1.9 & 1.9 & 1.9 \\
Minimum Particle Size ($\mu$m)&	0.5 & 10.3 & 22.4 & 71.0\\
\hline
\enddata
\vskip 0.05in
\scriptsize{Diameters and orbital distances from JPL/Horizons; density values from \citet{erosDensity,itokawaDensity}; $^a$assumed value}
\label{tab:examples}
\end{deluxetable}

\begin{deluxetable}{cccccccc}
\tablenum{2}
\tabletypesize{\footnotesize}
\tablecaption{Asteroid Observations}
\tablewidth{0pt}
\tablehead{
\colhead{Asteroid}   &
\colhead{UT Obs Date}   &
\colhead{V mag}   &
\colhead{T$_{exp}$ (sec)}   &
\colhead{n$_{exp}$}   &
\colhead{$\alpha$}  &
\colhead{Linear $\%~$Pol$^a$}  &
\colhead{$\theta_p$}  
}
\startdata
2006 SZ$_{217}$ & 2008-11-18 & 15.6 & 300 & 18 & $19.8^\circ$ & $0.1 \pm 0.2$ & $9 \pm 10$ \\
 & 2008-12-3 & 16.0 & 240 & 12 & $21.7^\circ$ & $0.3 \pm 0.1$ & $11 \pm 7$ \\
 & 2008-12-23 & 17.4 & 300 & 12 & $36.9^\circ$ & $1.4 \pm 0.5$ & $176 \pm 15$ \\
162900 & 2008-11-18 & 14.9 & 300 & 12 & $5.3^\circ$ & $-0.5 \pm 0.1$ & $86 \pm 13$ \\
 & 2008-12-3 & 14.9 & 120 & 12 & $16.4^\circ$ & $-0.2 \pm 0.1$ & $90 \pm 9$ \\
 & 2008-12-23 & 15.1 & 100 & 18 & $30.7^\circ$ & $1.1 \pm 0.1$ & $177 \pm 7$ \\
142348 & 2008-11-18 & 16.2 & 300 & 6 & $39.3^\circ$ & $5.1 \pm 0.5$ & $177 \pm 4$ \\
 & 2008-12-3 & 16.1 & 270 & 18 & $32.6^\circ$ & $2.4 \pm 0.1$ & $2 \pm 2$ \\
 & 2008-12-23 & 16.1 & 270 & 6 & $17.1^\circ$ & $0.1 \pm 0.1$ & $20 \pm 11$ \\
85236 & 2008-12-3 & 16.3 & 300 & 18 & $60.3^\circ$ & $3.8 \pm 0.1$ & $179 \pm 1$ \\
 & 2008-12-23 & 17.7 & 300 & 12 & $46.6^\circ$ & $2.5 \pm 0.5$ & $177 \pm 9$ \\
\hline
\enddata
\vskip 0.05in
\scriptsize{$^a~$quoted errors are $1\sigma$ statistical errors; systematic errors are $\approx0.05\%$.}
\label{tab.obs}
\end{deluxetable}

\newpage

\begin{center}
\begin{deluxetable}{lccccccccc}
\tablenum{3}
\tabletypesize{\footnotesize}
\rotate
\tablecaption{Derived Physical and Dust Properties}
\tablewidth{0pt}
\tablehead{
\colhead{Asteroid}   &
\colhead{$\alpha_0$ (deg)}   &
\colhead{h}   &
\colhead{q (AU)}  &
\colhead{H$_V$}  &
\colhead{$p_v$} & 
\colhead{D (km)} & 
\colhead{$n$} & 
\colhead{$d_{min}$ ($\mu$m)} & 
\colhead{$d_{pred}$ ($\mu$m)} 
}
\startdata
2006 SZ$_{217}$ & $\sim18$ & $0.08 \pm 0.03$ & 1.200 & 17.4 & $0.3^{+0.2}_{-0.1}$ & $0.8 \pm 0.2$ & $\sim1.64$ & $5^{+6}_{-2}$ & 8---32 \\
162900 & $19 \pm 3$ & $0.09 \pm 0.01$ & 1.215 & 15.8 & $0.24 \pm 0.03$ & $1.9 \pm 0.1$ & $1.7 \pm 0.1$ & $4.2^{+0.9}_{-0.7}$ & 3---13  \\
142348 & $17 \pm 2$ & $0.15 \pm 0.01$ & 1.119 & 18.2 & $0.14 \pm 0.01$ & $0.81^{+0.03}_{-0.02}$ & $1.60 \pm 0.06$ & $2.6 \pm 0.2$ & 8---32\\
85236 & $\sim18$ & $0.09 \pm 0.04$ & 0.850 & 18.5 & $0.2^{+0.2}_{-0.1}$ & $0.6 \pm 0.2$ & $\sim1.65$ & $4^{+5}_{-2}$ & 20---79  \\
Itokawa & $20 \pm 0.2$$^a$ & $0.084 \pm 0.001$$^a$ & 0.953 & 19.2 & $0.265 \pm 0.002$ & $0.373 \pm 0.001$$^c$ & $1.72 \pm 0.01$ & $4.41^{+0.03}_{-0.05}$ & 62.8\\
Eros & $21.7 \pm 0.9$$^b$ & $0.107 \pm 0.002$$^b$ & 1.133 & 11.2 & $0.203 \pm 0.004$ & $17.0^{+0.1}_{-0.2}$$^c$ & $1.79 \pm 0.03$ & $3.3 \pm 0.1$ & 0.5\\
\hline
\enddata
\vskip 0.05in
\scriptsize{$^a~$data from \citet{cellinoItokawa}.  $^b~$data from \citet{zgEros}.  $^c$calculated from polarimetry, thus differing slightly from {\it in situ} values.}
\label{tab.dust}
\end{deluxetable}
\end{center}

\clearpage

%\begin{figure}
%\centering
%\includegraphics[angle=-90,scale=0.6]{Figs/dmin.epsi}
%\protect\caption{
%The same as Fig~\ref{fig.dmax} but for various values of minimum grain size.
%}
%\label{fig.dmin}
%\end{figure}

%\begin{figure}
%\centering
%\includegraphics[angle=-90,scale=0.6]{Figs/index.epsi}
%\protect\caption{
%The same as Fig~\ref{fig.dmax} but for various values of index of refraction.
%}
%\label{fig.index}
%\end{figure}

%\begin{figure}
%\centering
%\includegraphics[angle=-90,scale=0.6]{Figs/power.epsi}
%\protect\caption{
%The same as Fig~\ref{fig.dmax} but for various values of power of the size distribution.
%}
%\label{fig.power}
%\end{figure}

\end{document}